\documentclass[
reprint,
superscriptaddress,
amsmath,
amssymb,
aps,
prl
]{revtex4-2}

\usepackage{comment}
\usepackage{graphicx}% Include figure files
\usepackage{dcolumn}% Align table columns on decimal point
\usepackage{bm}% bold math
\usepackage{lipsum}%generate dummy text
\usepackage{layouts}
\usepackage{physics}
\usepackage{amsmath,amsfonts,amsthm}

\usepackage[dvipsnames]{xcolor} 

\usepackage[colorlinks, linkcolor=NavyBlue, citecolor=NavyBlue, urlcolor=NavyBlue, breaklinks=true]{hyperref}

\newcommand{\pder}[2][]{\frac{\partial#1}{\partial#2}}

\usepackage{soul} %for highlighting text

\begin{document}

\title{
%Experimental realization of an inductively shunted transmon \\ with simultaneous charge and flux noise suppression}
%Simultaneously suppressing charge and flux noise sensitivity in an inductively shunted transmon with stable fluxon states} 
A superconducting qubit with noise-insensitive plasmon levels and decay-protected fluxon states} 
% Observation of long-lived fluxon states in an inductively shunted transmon qubit based on charge and flux noise insensitive plasmon states
% Observation of long-lived fluxon states in an inductively shunted transmon qubit that exhibits simultaneous charge and flux noise insensitivity
%Realization of simultaneous charge and flux noise insensitivity in an inductively shunted transmon that exhibits stable fluxon states
%Observation of long-lived fluxon states in a charge and flux noise insensitive inductively shunted transmon qubit
%Simultaneously suppressing charge and flux noise sensitivity in an inductively shunted transmon with stable fluxon states
\author{F. Hassani}
\affiliation{Institute of Science and Technology Austria, 3400 Klosterneuburg, Austria}
\author{M. Peruzzo}
\affiliation{Institute of Science and Technology Austria, 3400 Klosterneuburg, Austria}
\author{L.~N.~Kapoor}
\affiliation{Institute of Science and Technology Austria, 3400 Klosterneuburg, Austria}
\author{A. Trioni}
\affiliation{Institute of Science and Technology Austria, 3400 Klosterneuburg, Austria}
\author{M. Zemlicka}
\affiliation{Institute of Science and Technology Austria, 3400 Klosterneuburg, Austria}
\author{J.~M.~Fink}
\email{jfink@ist.ac.at}
\affiliation{Institute of Science and Technology Austria, 3400 Klosterneuburg, Austria}

\date{\today}

%cite these:
%Lamba system heavy fluxonium qubit: https://journals.aps.org/prl/abstract/10.1103/PhysRevLett.120.150504
%also
%Lambda transitions: https://journals.aps.org/prapplied/abstract/10.1103/PhysRevApplied.9.054046
% this is the other heavy fluxonium with flux control: https://journals.aps.org/prx/pdf/10.1103/PhysRevX.11.011010
% there should be one more from Manucharyan -  no?

% this paper has some good old references about macroscopic quantum tunneling: https://journals.aps.org/prl/abstract/10.1103/PhysRevLett.104.150405

% interesting thesis about heavy fluxonium: from dave group - find it again!

% relevant?? theory of the Ej>>(EC>>)EL limit https://journals.aps.org/prb/pdf/10.1103/PhysRevB.87.144518

% a theory paper about how to deal with inductive shunts: https://journals.aps.org/prb/abstract/10.1103/PhysRevB.94.144507

% the fluxonium molecule: https://journals.aps.org/prx/abstract/10.1103/PhysRevX.7.031037

%IST abstracts, chronological:
%https://meetings.aps.org/Meeting/MAR18/Session/L39.11
%https://meetings.aps.org/Meeting/MAR19/Session/E26.9
%https://meetings.aps.org/Meeting/MAR19/Session/E26.11
%https://meetings.aps.org/Meeting/MAR21/Session/X31.10

%list the relevant papers from Koch: the charge noise one but there is also Dempster et al, and more...

\begin{abstract}
The inductively shunted transmon (IST) is a superconducting qubit with exponentially suppressed fluxon transitions and a plasmon spectrum approximating that of the transmon. It shares many characteristics with the transmon but offers charge offset insensitivity for all levels and precise flux tunability with quadratic flux noise suppression. In this work we propose and realize IST qubits deep in the transmon limit where the large geometric inductance acts as a mere perturbation.
With a flux dispersion of only 5.1\,MHz we reach the 'sweet-spot everywhere' regime of a SQUID device with a stable coherence time over a full flux quantum. Close to the flux degeneracy point the device reveals tunneling physics between the two quasi-degenerate ground states with typical observed lifetimes on the order of minutes. In the future, this qubit regime 
%a smaller inductor 
could be used to avoid leakage to unconfined transmon states in high-power read-out or driven-dissipative bosonic qubit realizations. Moreover, the combination of well controllable plasmon transitions together with stable fluxon states in a single device might offer a way forward towards improved qubit encoding schemes. 
\end{abstract}

\maketitle
\section{Introduction}
%During the last two decades 
Since the first observation of coherent Rabi oscillations 
%in superconducting qubits 
two decades ago \cite{nakamura_coherent_1999,martinis_rabi_2002}, 
%reducing qubit decoherence still plays an essential role in developing a superconducting quantum processor. During these two decades, 
superconducting qubit coherence times
%times of the superconducting qubits 
improved by several orders of magnitude \cite{Siddiqi2021} - recently reaching the millisecond mark \cite{kjaergaard_superconducting_2020,somoroff_millisecond_2021}. The community owes this success to continuous parallel innovations and effort put into improving the fabrication process \cite{place_new_2021,somoroff_millisecond_2021,wang_transmon_2021}, more thorough shielding and isolation from the environment \cite{vepsalainen_impact_2020}, but also to a much improved understanding and control of the circuit sensitivities to various noise sources. 

Controlling the circuit potential and the resulting variance of the qubit state wave functions 
%quantum fluctuation 
provides an essential tool for reducing the dispersion of the qubit levels and for engineering noise protected states \cite{Gyenis2021}. 
%by delocalizing the wave function. More precisely, as the wave function is composed of a superposition of more charge or flux basis states (delocalizes), the dispersion of the qubit level becomes less sensitive to charge or flux noise respectively. 
%\cite{brooks_protected_2013,kitaev_protected_2006}. 
This strategy was particularly successful in the case of the transmon qubit, a charge qubit which operates in the limit of large Josephson to charging energy ratio $E_J/E_C\gg1$ \cite{koch_charge-insensitive_2007,Schreier2008}, thus delocalizing the qubit wavefunctions over multiple charge basis states and flattening its charge dispersion. More recently this was also achieved in the case of rf-SQUID type qubits by realizing quasi-charge qubits that operate in the challenging to realize high impedance, i.e.~low inductive to charging energy ratio  $E_L/E_C\ll1$ \cite{pechenezhskiy_superconducting_2020,peruzzo_geometric_2021}, thus delocalizing the wave function in phase and flattening the flux dispersion. 

\begin{figure}[t]
\centering
\includegraphics[width=0.85\columnwidth]{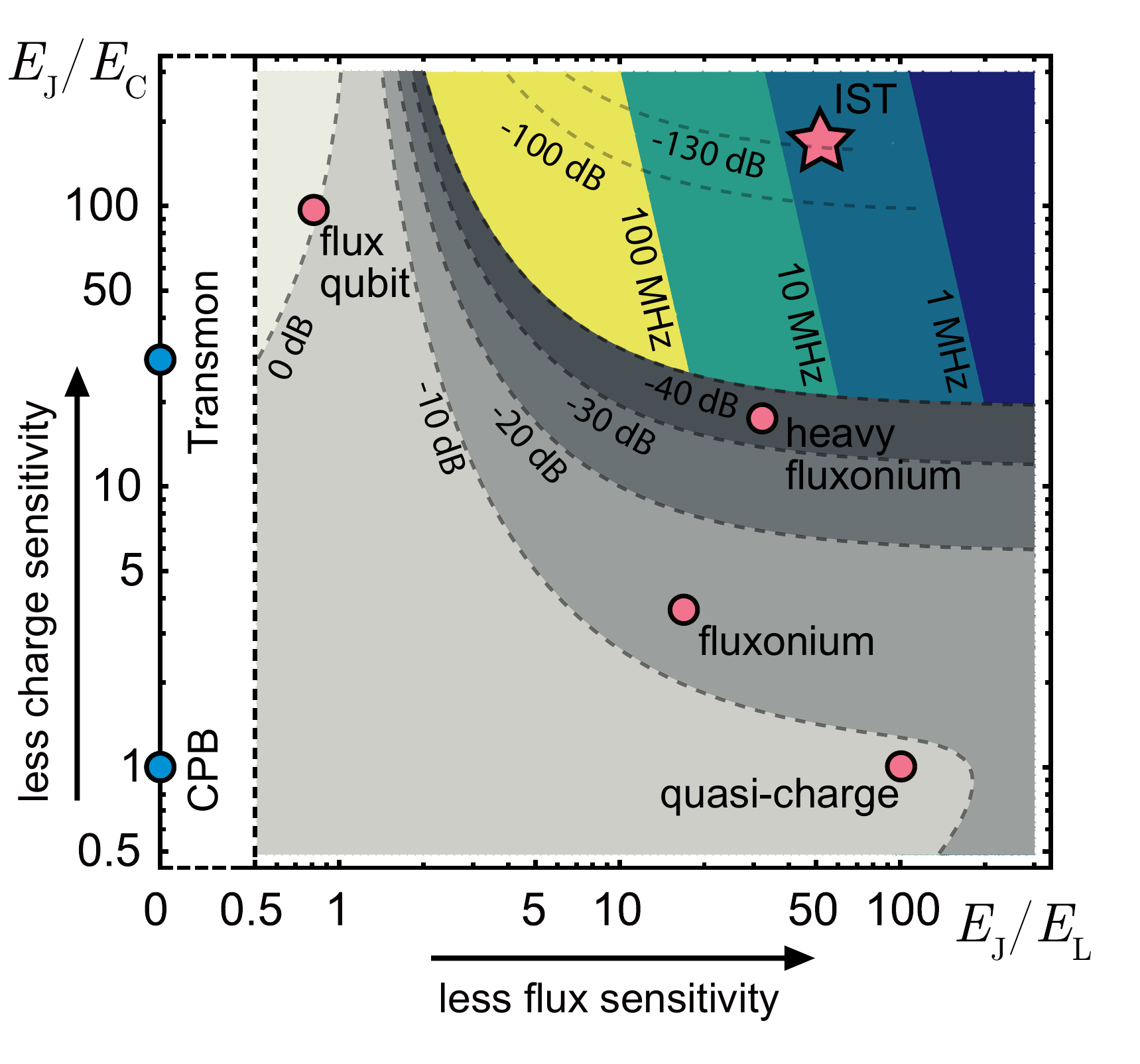}
\caption{
\textbf{Classification scheme for superconducting qubits.} We use the characteristic energy ratios $E_J/E_L$ and $E_J/E_C$ to parametrize the plot \cite{Girvin2004,Wendin2017,vool2017}. The values for the Cooper pair box (CPB), transmon, flux qubit, fluxonium, heavy fluxonium and quasi-charge qubit are taken from Refs.~\cite{bouchiat_quantum_1998,Schreier2008,yan_flux_2016,manucharyan_fluxonium_2009,Earnest2018,pechenezhskiy_superconducting_2020} respectively. The star depicts the parameters of the IST qubit (device A listed in Tab.~\ref{tab:summary}). The gray color scale and dashed gray lines are a contour plot of the matrix element of the lowest fluxon transition calculated at half flux $\varphi_\text{ext}=\pi$ on a logarithmic scale for the fixed Josephson energy $E_J/h = 29.93$\,GHz of device A. 
%The dashed gray line marks a value of $10^{-3}$ where fluxon transitions start to become inaccessible. 
The color coded contour areas in the fluxon transition suppressed region shows the flux dispersion of the plasmon state over a full flux quantum as calculated from Eq.~\ref{eq:disper} for the same $E_J$. 
%For small $E_C$, $E_L$ (large capacitance and inductance) compared to $E_J$ the band dispersion can reach less than $1$ MHz. 
} 
\label{fig:regime}
\end{figure}

\begin{figure*}[t]
\centering
\includegraphics[width=1\textwidth]{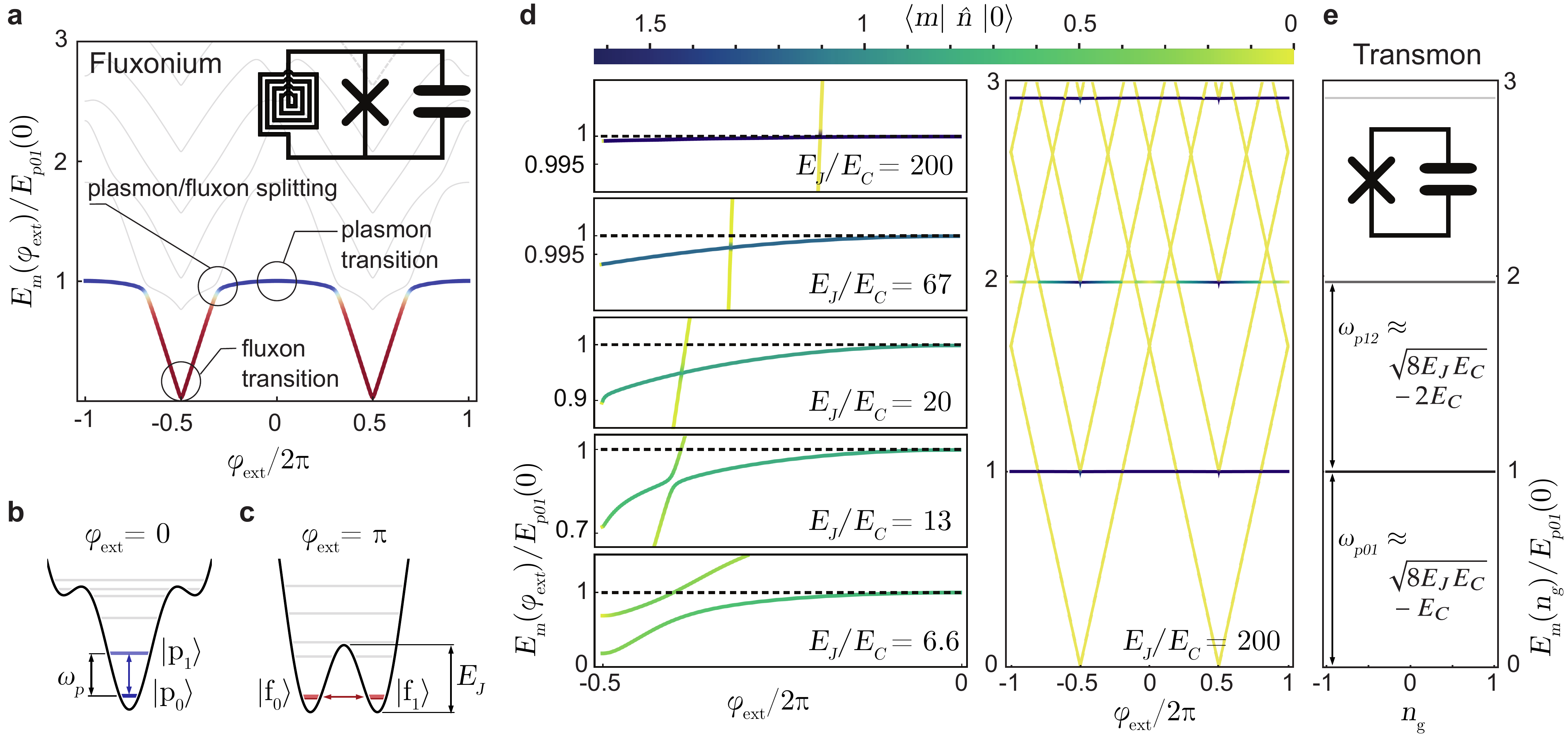}
\caption{
\textbf{Evolution from the fluxonium to the inductively shunted transmon spectrum.} 
\textbf{a}, Circuit (inset) and spectrum of a typical fluxonium device with $E_J/h=3$\,GHz, $E_L/h=0.5$\,GHz and $E_C/h=0.45$\,GHz \cite{nguyen_high-coherence_2019} as a function of external flux. 
%The colored line shows the ground to excited transition of the fluxonium and its dynamics with respect to applied external flux. 
The red and blue colors indicate flux and plasmon transition of the ground to first excited state respectively.
%, while the higher energy transitions are in dashed gray. 
%The inset shows the preliminary three element circuit of parallel Josephson junction, inductor and capacitance. b\&c) 
\textbf{b}, and \textbf{c}, show the potential for plasmon and fluxon transitions at zero and half flux respectively. 
\textbf{d}, First column shows the transformation of the fluxonium to the IST qubit spectrum by increasing the $E_J/E_C$ ratio ($E_C/h=150$\,MHz and $E_L/h=500$\,MHz), where the color scale indicates the calculated matrix elements. 
%As $E_J/E_C$ increases the flux transitions will be suppressed and only the plasmon states are accessible. 
The second column shows the full spectrum of the IST qubit including the diamond shaped flux levels with ultra-small matrix elements. 
%up the the third plasmon level. 
%The diamond shape transitions (yellow) are the suppressed flux transitions. 
The low dispersion plasmon levels are in agreement with the transmon spectrum shown in panel \textbf{e} for the same $E_J$ and $E_C$.  
%states however resemble the spectrum of the transmon, the only difference is that the spectrum of the IST qubit varies with flux instead charge offset. The third column on the right provides a comparison of a transmon spectrum with the same $E_J/E_C$ ratio and its circuit shown in the inset. 
} 
\label{fig: concept}
\end{figure*}

In this work we present a different strategy to achieve the latter, i.e.~the comparably easy to realize - but so far unexplored - limit of $E_J/E_C, E_J/E_L\gg1$ as shown in Fig.~\ref{fig:regime}. This limit does not rely on particularly high impedance but rather on making use of plasmon levels - a characteristic of charge qubits - in a rf-SQUID qubit geometry that traditionally relies on flux encoding. There are a number of proposals that introduce a variant of such a qubit as a suitable device to implement longitudinal 
%coupling instead of the usual transversal 
coupling \cite{richer_circuit_2016,richer_inductively_2017}, 
%to eliminate unwanted ZZ interactions, 
or to explore non-abelian many-body states \cite{Hafezi2014}. More moderate versions of it are being explored to optimize the transmon towards higher anharmonicity, controlled flux tunability and resulting higher gate fidelities \cite{liu2021}.

Even though the plasmon qubit encoding in the deep IST limit studied here shares many similarities with the transmon - including its eigenenergy, anharmonicity and transition matrix elements - there are also a number of important differences. The large inductive shunt of the IST decompactifies the phase of the transmon \cite{devoret2021} and localizes continuous qubit wave functions in wells with discrete flux number. The inductor therefore leads to a quadratic confinement of the qubit wavefunction in the phase variable and can be used to avoid dissipative leakage to neighboring wells that are not part of the computational basis of charge qubits with compact phase. 
%\cite{Venkatraman2019}. 
Such dynamical instabilities \cite{sank_measurement-induced_2016,verney_strongly_2018,lescanne_escape_2019} currently limit not only high-power qubit readouts but are also believed to prevent the generation of larger photon number cat-qubits that would be beneficial for improved $T_1$ protection \cite{grimm_stabilization_2020}. 
Another major difference to the transmon is that the higher energy states of the IST do not suffer from an increased charge dispersion that makes the transmon susceptible to charge noise and limits its use for resource efficient, higher-dimensional quantum information processing with qudits \cite{Ringbauer2021}. 
%Finally, the rf-SQUID geometry of the IST allows for controlled flux fine tuning of a single junction device with a transmon spectrum. 

After going over the theory of the IST qubit and its relationship to the fluxonium and transmon, we present the device design, spectroscopy and time-domain coherence measurements for 3 devices with different $E_L$. We furthermore show that the established tools from circuit QED, i.e.~the spectroscopically determined plasmon transition frequencies, can be used to read out the long-lived local fluxon ground states, which reveal  interesting tunneling physics away from zero flux and at elevated temperatures.

\begin{figure*}[t]
\centering
\includegraphics[width=1.85\columnwidth]{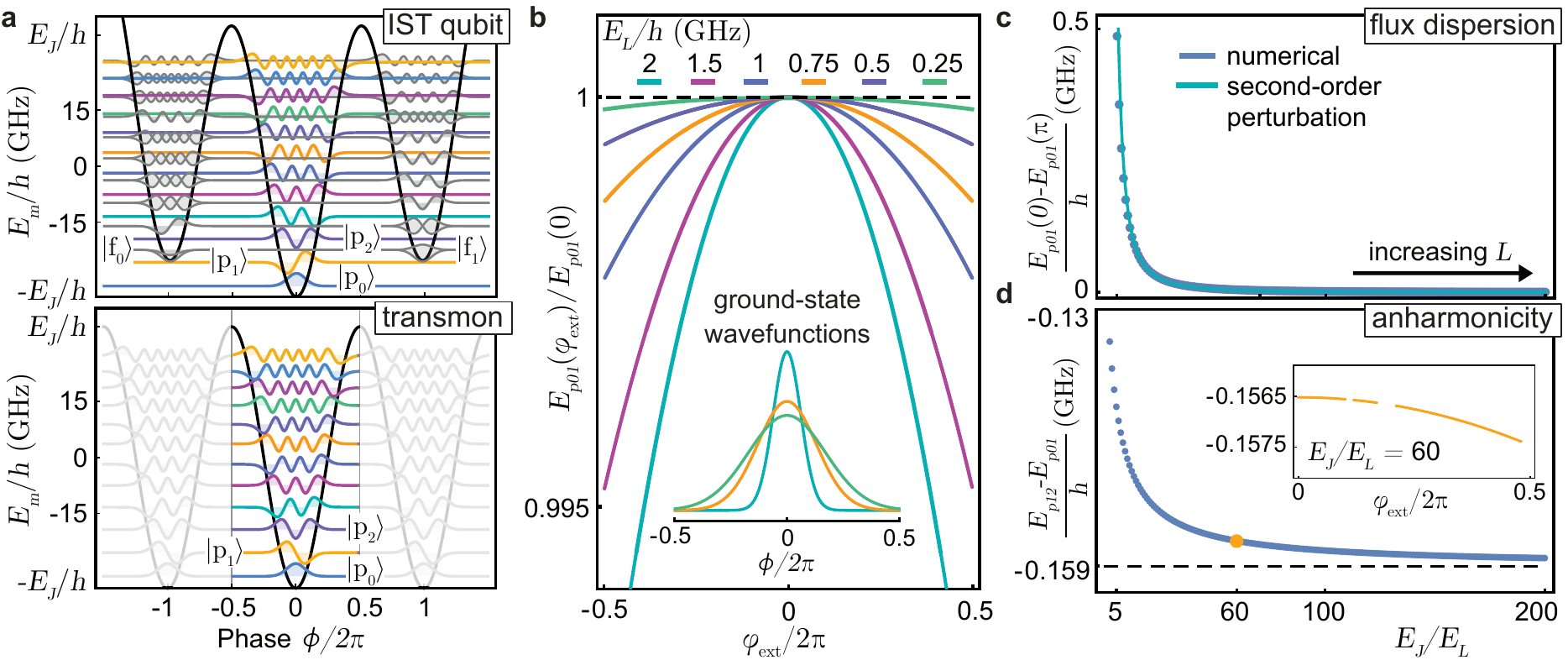}
\caption{
\textbf{Quantitative comparison of the IST and transmon qubits.}
\textbf{a}, The potential (black line), wave functions (colored and gray lines) and eigenenergies (y-axis offsets) of the IST qubit (top) for $\varphi_\text{ext}=0$ and the equivalent transmon without inductive shunt (bottom). The high tunneling barrier given by $E_J/h=35$\,GHz along with the heavy mass ($E_C/h=0.15$\,GHz) renders the quadratic confinement of the wells given by $E_L/h=0.5$\,GHz of the IST to be a mere perturbation for the lowest energy levels. This leads to plasmon wavefunctions (colored) resembling closely those of the transmon (bottom). 
%The transmon and IST plasmon state therefore form extremely similar wavefunctions that share the same properties. 
The gray wave functions in the top panel show the practically inaccessible flux transitions for (exactly) $\varphi_\text{ext}=0$. 
\textbf{b}, Lowest energy plasmon transition (colored lines) as a function of $\varphi_\text{ext}$ for the same fixed $E_J$ and $E_C$. The choice of $E_L$ determines the amount of phase confinement. Higher inductance leads to smaller phase confinement, a larger ground state wavefunction variance (inset) and a strongly reduced flux dependence.
 %of the lowest energy plasmon transition (colored lines).
% is an extra knob to control the qubit dispersion with respect to flux. The higher the inductance, the flatter the band. In the inset, the wave function of the ground state is shown. Higher inductance leads to higher characteristic impedance($Z_C=\sqrt{L/C}$), which delocalizes the wave function over the flux, making it more robust against flux noise. 
\textbf{c}, The calculated full flux dispersion as a function of $E_J/E_L$ for the same fixed $E_J$ and $E_C$. The results from numerical diagonalization (blue points obtained with Ref.~\cite{Groszkowski2021}) agree with the quadratic suppression predicted by Eq.~\ref{eq:disper} based on second-oder perturbation theory (green line).
%  in green and compares it with the results of the numerical diagonalization (using scQubits python library \cite{Groszkowski2021}) in blue.  The $E_J/E_L$ ratio suppresses the band dispersion quadratically. 
%The figure includes the calculated dispersion ($\nu_{ge}|_{\phi_{ext}=0}-\nu_{ge}|_{\phi_{ext}=\pi}$) 
\textbf{d}, The calculated qubit anharmonicity as a function of $E_J/E_L$ for the same fixed $E_J$ and $E_C$. In the high $E_J/E_L$ limit the anharmonicity of the IST (blue points) converge to that of the transmon (dashed line).
%\hl{ with a transition frequency of $\omega_{p12}/(2\pi)\approx6$\,GHz. }
%Both figures are for a fixed transition frequency of $6$ GHz with charging energy of $150$ MHz. 
The inset shows that the change of this anharmonicity as a function of external magnetic flux is very small 
%($\approx 600$\,kHz at $E_J/E_L=70$) 
for large values of $E_J/E_L$.} 
\label{fig: Theory calculations}
\end{figure*}

\section{Results}
\subsection{Theory}
%Looking back at the circuit of the fluxonium qubit, it is very tempting to identify it as a transmon (parallel circuit of a Josephson junction and a capacitance) being inductively shunted. However, the spectrum of a typical fluxonium devices shown in Fig.1a are nothing like a transmon. The reason for this becomes clearer by looking at the Hamiltonian of the circuit written as
The Hamiltonian of the IST qubit is that of the rf-SQUID shown in the inset of Fig.~\ref{fig: concept}a and given as
\begin{align}\label{eq:fluxonium_hamiltonian}
H= 4 E_C\hat{n}^2-E_J\cos{(\hat{\phi})}+\tfrac{1}{2}E_L(\hat{\phi}+\varphi_\text{ext})^2,
\end{align} 
%Where $E_C$ the charging energy, $E_L$ the inductive energy and $E_J$ the Josephson energy, are the essential energy scales governing the circuit. 
where the first two terms describe a regular transmon qubit with the two canonical variables 
%\hl{Cooper pair} 
charge $\hat{n}$ and phase $\hat{\phi}$. 
%where the fictions phase particle is confined by the cosine potential. 
Adding the inductive energy term adds a quadratic confinement in the flux degree of freedom and
%to the transmon Hamiltonian 
lifts the periodicity of the $\cos{(\hat{\phi})}$ potential which enables flux transitions between neighboring wells. 
%in the new potential landscape. 
The spectrum 
%of Eq.~\ref{eq:fluxonium_hamiltonian}, 
shown in Fig.~\ref{fig: concept}a, unlike the transmon, is invariant to charge offset \cite{koch_charging_2009} and is a function of the external magnetic flux $\varphi_\text{ext}=2\pi\Phi_\text{ext}/\Phi_0$ instead. 
%%%%%%%%%%%%%%%%%%%%%%%%%%%%%%%%%%%%%%%%%%
%%%%%%%%%%%%%%%%%%%%%%%%%%%%%%%%%%%%%%%%%%
%This is expected to be particularly valid in our experimental realization that does not rely on multiple terminals or islands \hl{\cite{ferguson2013,dipaolo2021}} to form a large inductance shunting the single small Josephson junction.
%%%%%%%%%%%%%%%%%%%%%%%%%%%%%%%%%%%%%%%%%%
%%%%%%%%%%%%%%%%%%%%%%%%%%%%%%%%%%%%%%%%%%
%In Fig.~\ref{fig: concept}a the dynamics of the first excitation level of a fluxonium is plotted. 

At zero external flux $\varphi_\text{ext}=0$, the first transition between the ground and first excited state is located within one well as shown in Fig.~\ref{fig: concept}b and approximately given by the plasmon frequency $\omega_p=\sqrt{8 E_J E_C}$. 
%Looking at the potential configuration shown in Fig.~\ref{fig: concept}b the transition is within one well. 
In this limit the plasmon transition energies are mostly a function of the shape of the bottom of the single cosine well and therefore comparably insensitive to external flux. 

As the magnetic field is tuned to half flux $\varphi_\text{ext}=\pi$ the potential changes to a double well configuration shown in Fig.~\ref{fig: concept}c favoring low energy fluxon transitions - the natural basis states of loop-type qubits such as flux and fluxonium qubits \cite{mooij_josephson_1999,manucharyan_fluxonium_2009,kwon2021}. In this bias condition long energy relaxation times \cite{lin_demonstration_2018-1} as well as protection from quasi-particle loss \cite{Pop2014} have been demonstrated.
%%%%%%%%%%%%%%%%%%%%%%%%%%%%%%%%%%%%%%%%%%
%%%%%%%%%%%%%%%%%%%%%%%%%%%%%%%%%%%%%%%%%%
%\cite{lin_demonstration_2018-1}. 
%%%%%%%%%%%%%%%%%%%%%%%%%%%%%%%%%%%%%%%%%%
%%%%%%%%%%%%%%%%%%%%%%%%%%%%%%%%%%%%%%%%%%
%Therefore, flux type devices are designed to work at half flux quantum \cite{mooij_josephson_1999,manucharyan_fluxonium_2009,pechenezhskiy_superconducting_2020-1,yan_flux_2016}. 

At intermediate flux values the the spectrum in Fig.~\ref{fig: concept}a exhibits a plasmon/fluxon transition splitting. In the high inductance limit $E_J/E_L\gg1$ the size of this splitting is a measure of the circuit's desire to favor phase slips or charge tunneling as shown in Fig.~\ref{fig: concept}d. In the low-capacitance (light) regime characterized by small $E_J/E_C \sim 1$ the splitting is very large. In the case of ultra-high impedance the splitting opens up and forms flat Bloch-bands that form the basis states of the recently realized quasi-charge qubit \cite{pechenezhskiy_superconducting_2020, peruzzo_geometric_2021}. In the high-capacitance (heavy) regime characterized by $E_J/E_C\gg1$ on the other hand the circuit is dominated by the plasmonic character. 

The diamond-shaped fluxon transitions in Fig.~\ref{fig: concept}d are exponentially suppressed with the matrix element calculated to be as low as $10^{-13}$ due to the heavy nature of device A with $E_J/E_C=182$ as shown in Fig.~\ref{fig: Theory calculations}. Here the plasmon/fluxon splitting is closed and the plasmon levels form flat bands with extremely small flux dispersion on the order of a few MHz, cf.~Fig.~\ref{fig: concept}d and dashed contour lines in Fig.~\ref{fig:regime}. Figure \ref{fig: concept}e shows that in this regime the IST plasmon transition energies are in good agreement with those of the equivalent transmon circuit without the large inductive shunt - except that those bands are flat with respect to the gate charge $n_g$ rather than $\varphi_\text{ext}$.

In Figure~\ref{fig: Theory calculations}a we compare the potential, eigenenergies and wave functions with those of the transmon to acquire more intuition about the properties and spectrum of the IST qubit. While the transmon potential and wavefunctions extend periodically from minus infinity to infinity, the lowest energy IST wave functions are localized in one specific well. Because the inductive confinement lifts the neighboring wells by a small energy compared to the depth of the wells given by $E_J$, the shape of the potential, the resulting plasmon wavefunctions, and eigenenergies resemble very closely those of the transmon. The matrix elements between flux states (gray wavefunctions) are exponentially suppressed by the very large barrier between the wells (compared to the plasmon energy). 

In the small $E_L$ (large $L$) limit the shape of the well is mostly determined by the Josephson energy and the lowest plasmon transition energy therefore becomes extremely flat vs. flux, as shown in Fig.~\ref{fig: Theory calculations}b, exhibiting a relative flux dispersion of less than one part in a thousand. As expected, this insensitivity is accompanied with an increased variance of the ground state wavefunction in phase as shown in the inset. The IST qubit therefore realizes flux noise insensitivity by increasing $E_J/E_L$ in analogy to the charge noise insensitivity of the transmon obtained for large $E_J/E_C$.
% realized protection against charge noise compared to the cooper pair box, 

%The subset of Fig. \ref{fig: Theory calculations}b shows the wave function of the ground state for different inductive energies. The spread of the wave function in flux basis with lower inductive energies leads to lower dispersion with external flux. 

%formed by the fluxon transition depicted in Fig.~\ref{fig: concept}d, we look at the potential, eigenenergies and wave functions of the IST qubit and compare it with transmon(Fig. \ref{fig: Theory calculations}a). 
%The plasmon energies flatten because the shape of the well is mostly determined by the Josephson energy and does not change much with external flux over the range of one flux quantum. 

%The effect of the inductive energy on the qubit level is shown in Fig. \ref{fig: Theory calculations}b where a zoom in on the first plasmon transition within one flux quantum is provided. The inductive energy provides a knob to further suppress the dispersion with respect to external flux. As in the transmon where $E_J/E_C$ brought protection against charge noise, the IST qubit displays flux noise suppression by increasing $E_J/E_L$. The subset of Fig. \ref{fig: Theory calculations}b shows the wave function of the ground state for different inductive energies. The spread of the wave function in flux basis with lower inductive energies leads to lower dispersion with external flux. 

To get more insight into the scaling of the flux noise protection we solve the Hamiltonian Eq.~\ref{eq:fluxonium_hamiltonian} under the assumption that the inductive part of the Hamiltonian acts as a local perturbation ($E_L\ll E_J$). The Materials and Methods section covers the derivation and a comparison to numerical results. In the limit where $E_J/E_C\gg1$ we obtain a simple expression for the flux dispersion of the first plasmon qubit transition 
 \begin{align}\label{eq:disper}
\pder[\omega_{p01}]{\varphi_\text{ext}}=-\frac{\sqrt{8E_J E_C}}{4\hbar(E_J/E_L)^2}\varphi_\text{ext}.
\end{align}   
It shows that the $E_J/E_L$ ratio provides a quadratic suppression of both, the first and second order derivative, which are the relevant quantities for qubit dephasing at intermediate and zero flux \cite{Ithier2005}. 
Furthermore Eq.~\ref{eq:disper} also shows that, in analogy to the Cooper pair box where the transition frequency is a function of charge offset squared $\omega_{01}(n_g^2)$, the IST qubit transition is also given by a parabola but versus external flux $\omega_{p01}(\Phi^2_\text{ext})$. 

Figure~\ref{fig: Theory calculations}c shows the full dispersion $E_{p01} (0) - E_{p01} (\pi)$, calculated with Eq.~\ref{eq:disper} for a fixed set of $E_J$ and $E_C$.
%ratio (or a fixed plasma frequency $\sqrt{8E_J E_C}$). 
The quadratic prediction (green line)
%of Eq. \ref{eq:disper} 
matches very well with the numerical results (blue points) for a large range of $E_J/E_L$. Panel d shows the calculated anharmonicity $E_{p12} - E_{p01}$ of the qubit for the same parameters with $\omega_{p01}\approx 6$\,GHz as a function of the $E_J/E_L$ ratio. As the inductance of the superinductor increases the Hamiltonian in Eq.~\ref{eq:fluxonium_hamiltonian} converges to that of the transmon and therefore the anharmonicity of the qubit also converges to the transmon anharmonicity (dashed line).
%approximately $-E_C$ (dashed line). 
However, for low inductance the parabolic potential dominates which results in lower anharmonicity.  

%Finally, the level spacing of the plasmon transition imitates transmon since the inductive energy is now only a perturbation to transmon Hamiltonian and both the eigenenergies and eigenfunctions look very similar to transmon within one well. To specify, in Fig. \ref{fig: Theory calculations}a the two lowest states laying at the bottom of the well (denoted by the $\omega_{ge}$) form the IST qubit. 

\subsection{Experimental realization}
\begin{figure}[t]
\centering
\includegraphics[width=1.0\columnwidth]{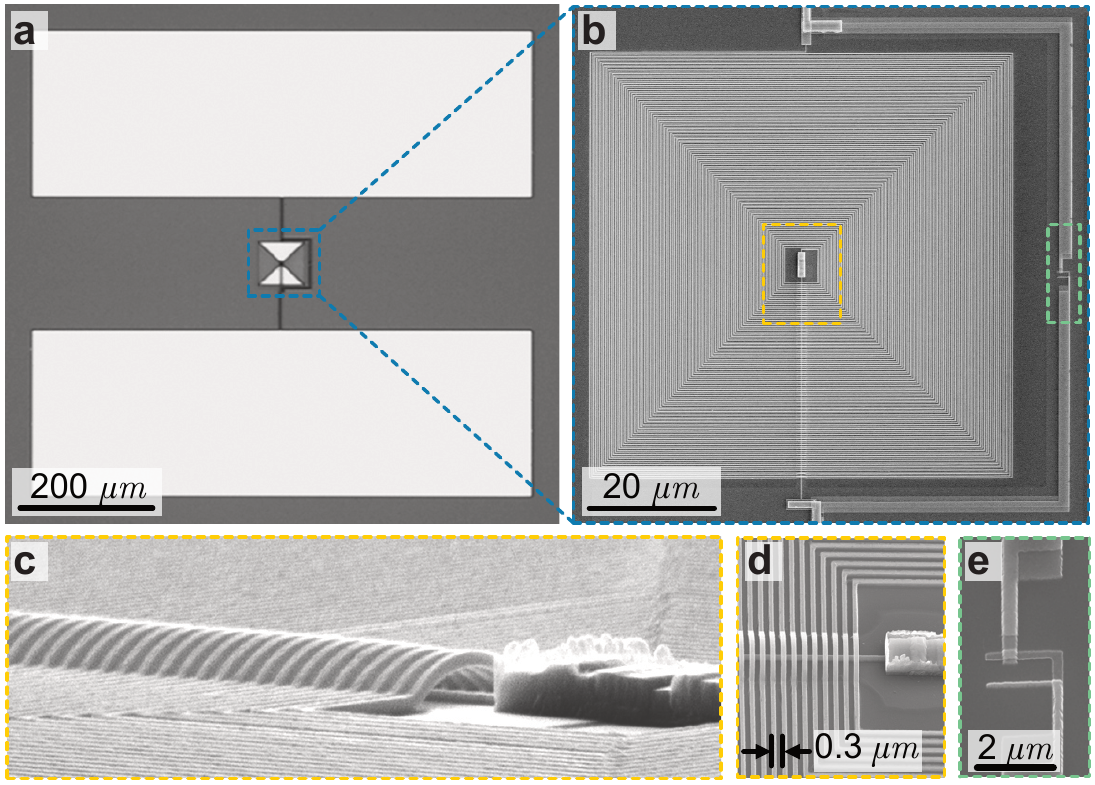}
\caption{
\textbf{IST qubit scanning electron microscope images.} 
\textbf{a}, Overview image of the aluminum capacitor pads (white) fabricated with an inductively coupled plasma etching recipe on high resistivity silicon (dark gray).
\textbf{b}, Enlarged view of the aluminum coil inductor and Josephson junction shunting the qubit capacitor. 
\textbf{c}, Isometric view of the central airbridge part of the inductor and the patch layer, which is deposited after ion gun etching to ensure a reliable electrical contact between the coil, capacitor and junction aluminum layers. 
\textbf{d}, Enlarged view of the center of the coil with 99 turns and a wire width and spacing of 150 nm respectively. It is fabricated using an inductively coupled plasma etching recipe and device C from the first generation exhibits a few shorts which leads to a three times lower inductance. 
\textbf{e}, Enlarged view of the Josephson junction fabricated with the Dolan bridge method \cite{dolan_offset_1977}.
}
%The device image including scanning electron microscope (SEM) images of the geometric superinductor and the Josephson junction. The device was fabricated on high resistivity silicon. The Josephson junction was fabricated through the well known shadow evaporation technique while its area and oxidation time were adjusted to give $E_J=35$ GHz approximately. The antenna pads were fabricated with normal liftoff techniques while the geometric superinductor was dry etched to increase the yield of the device. The antenna pads ensure a charging energy of about $200$ MHz and coupling to 3D cavity of about $100$ MHz. The effective number of turns in the superinductor control the inductive energy between $0.6$ to $1.6$ GHZ. The $300$ nm pitch size guaranties that the parasitic modes of the coil will not interfere with the design and readout of the qubit\cite{peruzzo_geometric_2021,peruzzo_surpassing_2020}.} 
\label{fig:device}
\end{figure}

The three studied IST qubit devices are based on the 3D transmon design \cite{paik_observation_2011} with a single Josephson junction with $E_J/h\approx 30$\,GHz
%To reach the limit of IST qubit, we control the area and oxidation time of the Josephson junction to obtain Josephson energies of the order $30$ GHz. Then, using the transmon 3D design \cite{paik_observation_2011} 
and a shunting capacitance of $C_s\approx100$\,fF as shown in Fig.~\ref{fig:device}. The large inductor shunting the Josephson junction is based on a 
%geometric superinductance \cite{peruzzo_surpassing_2020}, a 
miniaturized planar coil \cite{peruzzo_surpassing_2020} with a large inductance of $100 - 300$\,nH.
%varying number of turns producing inductive energy of 
The effective qubit parameters are listed in Tab.~\ref{tab:summary} and the fabrication details are to be found in the Materials and Methods section. 
%The fabricated device is shown in Fig. \ref{fig:device}, showing the independent parts of the circuit fabricated on high resistivity silicon chip. First, the capacitive pads of the device along with the cross wire of the coil are written with electron beam lithography and normal liftoff processes. Then, a resist covers the cross wire and is reflowed to shape the bridge crossovers. This is followed by an aluminum evaporation and dry etching to form the geometric superinductor. In the next step, the Josephson junction is added using standard shadow evaporation technique\cite{dolan_offset_1977}. Finally, all layers are connected with one another using an ion milling procedure followed by a deposition of a thick layer of aluminum to ensure connectivity. The detailed recipe of the fabrication is discussed in Appendix C. 

\begin{figure*}[t]
\centering
\includegraphics[width=2\columnwidth]{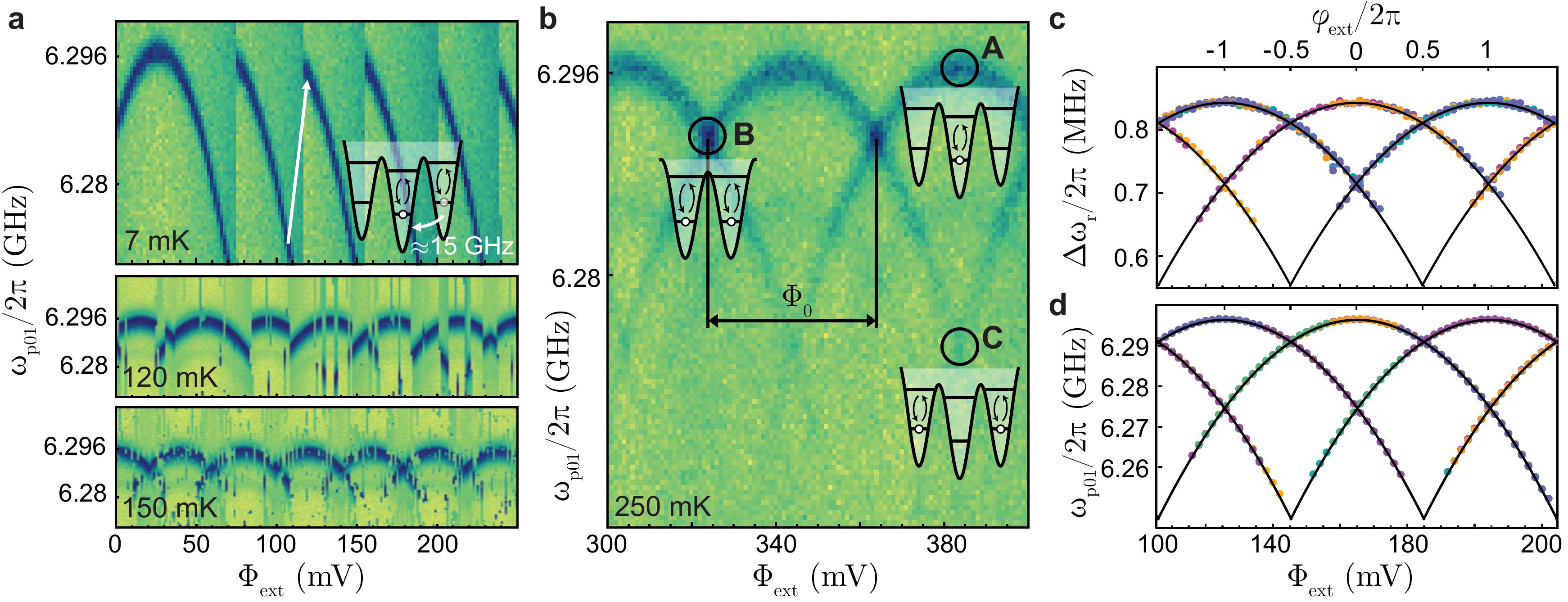}
\caption{
\textbf{Qubit spectroscopy and macroscopic quantum tunneling.}
\textbf{a}, Measured spectrum of the lowest energy plasmon mode versus magnetic flux in units of the voltage applied to an external bias coil using device B. At the base temperature the parabolic qubit spectrum exhibits discontinuities when a certain normalized bias value is exceeded. Each jump to higher frequencies corresponds to a quantum tunneling event of the circuit's fictitious phase particle that remains trapped in a stable flux state up to and beyond a full flux quantum of external flux corresponding to $\approx 15$\,GHz above the ground state \cite{Leggett1980,Martinis1985,martinis2020}. This process is shown in the inset with arrows (not to scale). At higher temperatures these tunneling events are triggered by thermal fluctuations and become more frequent - also for small bias values. 
\textbf{b}, Plasmon level spectroscopy at 250\,mK results in a smooth spectrum that contains a thermal mixture of all tunneling events and populations. This allows to identify the magnetic flux quantum $\Phi_0$ and individual parabola that correspond to specific and distinguishable potential wells with an integer flux $m$ being occupied. Point \textbf{A} of the inset labels refers to the plasmon transition sweet spot where $\Phi_\text{ext}=m\Phi_0$ and, starting from the flux ground state, the highest transition frequency is measured. At the half flux point \textbf{B} the two neighboring wells become degenerate with identical and somewhat lower plasmon transition frequency. Point \textbf{C} identifies the degeneracy point between two wells left and right of the global ground state. 
\textbf{c}, and \textbf{d}, show a fit to the plasmon qubit and readout resonator dispersive shift $\omega_r/2\pi-10.459$~GHz, obtained at base temperature and based on multiple flux sweeps and data sets (color coded). 
%The measured resonator dispersive shift $\Delta\omega_r/2\pi=\omega_r/2\pi-10.459$~GHz. 
The fit (black lines) was obtained by first fixing $E_J$ and $E_C$ using point \textbf{A} in panel b, then using point \textbf{B} and \textbf{C} to obtain the inductive energy $E_L$.}
\label{fig:spec}
\end{figure*}

\begin{table}[t]
\caption{
\textbf{Extracted qubit parameters.
% for devices A-C. 
} 
The reported coherence times are measured at $\varphi_\text{ext}=0$. $\delta\nu_{p01}$ refers to the measured qubit dispersion over the full flux range.}
\small\centering
\label{tab:summary}
\begin{tabular}{||c|c|c|c|c|c|c|c|c||} 
\hline
 & $E_J/h $ & $E_C/h $ & $E_L/h $  & $g_0/(2\pi)  $ & $T_1$  & $T_2$ & $\nu_{p01}$ & $\delta\nu_{p01}$ \\
 & (GHz) & (GHz) & (GHz) & (MHz) & ($\mu$s) & ($\mu$s) & (GHz) & (MHz) \\
\hline
\hline
\textbf{A} & 29.93 & 0.164 & 0.56 & 107.7 & 15.5 & 13.0 & 6.122 & 5.1 \\
\hline
\textbf{B} & 31.13 & 0.165 & 0.56 & 119.6 & 21.0 & 27.8 & 6.296 & 5.6 \\
\hline
\textbf{C} & 33.34 & 0.170 & 1.60 & 86.3 & 17.4 & 22.6 & 6.720 & 40.0 \\
\hline
\end{tabular}
\end{table}

The fabricated devices are packaged in a rectangular 3D cavity made from oxygen free copper with the first resonance mode $\nu_{r}\approx10.48$\,GHz, an internal quality factor of $Q_i=2.7\times 10^4$ and a total loss rate of $\kappa \approx 1$\,MHz. 
%The qubit-cavity coupling of $g_0=90$\,MHz is determined by the chosen capacitor geometry and cavity width of $9$\,mm.
%for control and readout. 
The cavity is then attached to the cold plate of a dilution refrigerator at a temperature of $7$ mK. The qubit is controlled and read out via the cavity port using microwave pulses passing through multiple stage of attenuation, a $12$ GHz lowpass filter, an Eccosorb filter and finally a circulator to reach the cavity. The qubit readout is done based on the reflected signal that passes through two stages of isolators, a $8-12$ GHz band pass filter, a low-noise high electron mobility amplifier at the $3$\,K stage followed by another low-noise amplifier and demodulation at room temperature. We use a large radiation shield that is coated with a mixture of Stycast and carbon powder and thermalized at the mixing chamber. 
% to ensure the high frequency radiation that might reach the sample is absorbed. 
Inside it, the cavity is located on the bottom part of a double layer cryogenic $\mu$-metal shield to minimize stray magnetic fields. 
%acting on the qubit's loop. The setup chain and its microwave components are discussed in more detail in \cite{peruzzo_surpassing_2020}.

\begin{figure*}[t]
\centering
\includegraphics[width=1.5\columnwidth]{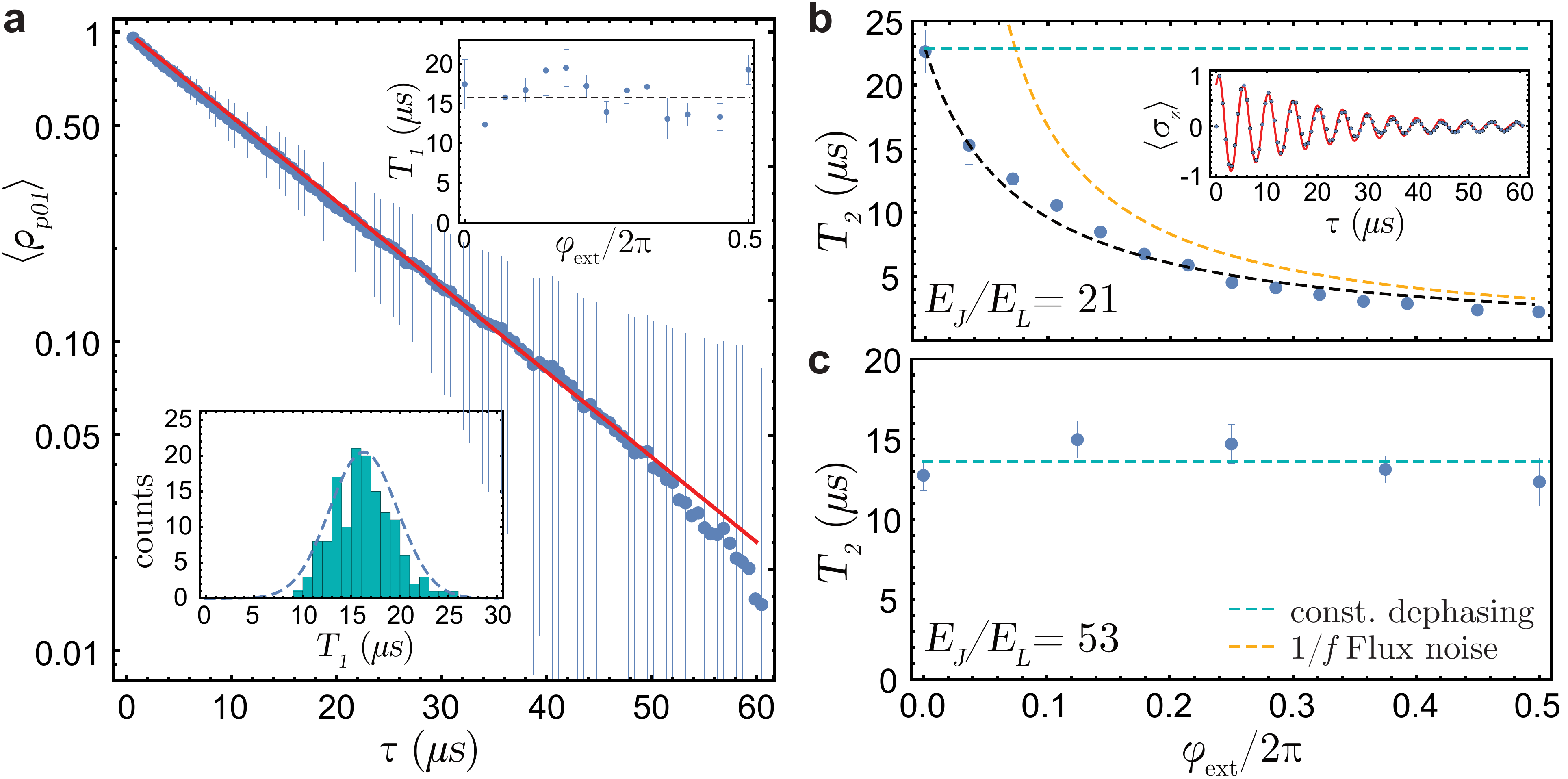}
\caption{\textbf{Qubit plasmon relaxation and dephasing measurements.}
\textbf{a}, The normalized readout voltage proportional to the average excited state population of device C after a $\pi$ pulse excitation of the $| p_{01}\rangle$ state at various flux values. %\hl{at $\varphi_\text{ext}=0$}. 
The blue points are the mean of all measured traces and the bars show the standard deviation of the individual measurements. The data does not show a direct indication for quasi-particle induced loss and fits well to a single decay exponential function (red line) yielding $T_1=15.8\,\mu$s. %indicates no quasi-particle density tunneling over the single Josephson junction. 
% measured for device $A$. %Here, $\tau$ is the delay between a calibrated $\pi$ pulse and the measurement pulse probing the resonator. 
%The top right inset shows the $T_1$ values extracted from the same data sets in the same way but as a function of  external flux. 
The histogram of all 120 measured relaxation times (bottom inset) agrees with a single-peaked Gaussian envelope (dashed line). 
The measured $T_1$ times are approximately constant vs.~external flux (top inset, error bars show statistical standard error) but the observed fluctuations around the mean (dashed line) suggest two-level-system coupling and dielectric losses.
%, as reported in \cite{burnett_decoherence_2019}.
% of the $T_1$ distribution supports the argument of $T_1$ being not dependant on external flux. 
\textbf{b}, and \textbf{c}, show $T_2$ decoherence times obtained from standard Ramsey measurements (top inset) for device C with $E_J/E_L=21$ and device A with $E_J/E_L=53$, respectively. 
%measured decoherence time 
%The full flux noise protection scheme at IST qubit for two sample devices, with one having an inductive energy of $1.6$ GHz(weak protection) and the other with $E_L=0.6$ GHz(strong protection). 
The maximum $T_2=22.6\,\mu$s of device C 
% ke=363 kHz ki=446 kHz
is strongly reduced by flux noise away from the integer flux sweet spot. We fit the flux dependence (dashed black line) using the measured mean $T_1$,
%\hl{$T_1=15.8\,\mu$s},
%max \hl{$T_1=17.4\,\mu$s} 
a thermal photon shot noise in the resonator $T_\text{th}\approx108$\,mK (dashed cyan line) 
%n_th=0.008 device C
%n_th=0.028 device A
and a $1/f$ flux noise amplitude of $A_\Phi=98\,\mu\Phi_0$ (dashed yellow line). 
% which taken together (dashed black line) agrees with the measured decay (blue points). 
Device A shown in panel \textbf{c}
% ke=618 kHz ki=361 kHz
on the other hand exhibits strong dephasing protection due to its large inductance. Over the full flux range $T_2$ is scattered around the mean $T_2=13.6\,\mu$s (dashed cyan line) without a clear flux dependence. 
It was possible to measure the $T_1$ and $T_2$ data at each flux value for around 30 minutes without unwanted switching events, also for values very close to half flux. 
%corresponds to a single particular potential well and near the degeneracy point (point \textbf{B} in Fig~\ref{fig:spec}b), where the probability of thermally activated tunneling increases, it was still possible to repeat the relaxation and dephasing measurements to acquire plotted means and standard deviations over the course of around 30 minutes without unwanted switching events.
%is dominated by the mean $T_1=15.5\,\mu$s together with a fitted \hl{$T_r\approx140$\,mK} (dashed cyan line).
}  
\label{fig:coherence}
\end{figure*}

We use dispersive readout and two tone spectroscopy \cite{wallraff_strong_2004} to obtain the qubit spectrum shown in Fig.~\ref{fig:spec}a. 
%shows the first excitation level of device B at $\nu_{ge}\approx\sqrt{8E_J E_C}-E_C$. 
Surprisingly, the spectrum at the base temperature of the dilution refrigerator does not show a periodic behavior with external flux as predicted by %the Hamiltonian in 
Eq.~\ref{eq:fluxonium_hamiltonian}. Flux periodicity is a crucial feature of any flux-tunable device 
%such as fluxonium or flux qubit, since 
as it provides the unit-less flux 
%fitting anchor point 
scaling and in turn allows to infer the qubit energies. 
%\cite{yan_flux_2016,manucharyan_fluxonium_2009,pechenezhskiy_superconducting_2020-1}. 
In usual rf-SQUID devices - including very heavy fluxonium qubits \cite{Earnest2018, lin_demonstration_2018, Vool2018, Zhang2021} - the global ground state of the system switches from one well to a neighboring well of the potential landscape at $\varphi_\text{ext}=\pi$ as a function of external flux and the fictitious phase particle always moves to the ground state well due to the non-negligible inter-well coupling.
%\cite{lin_demonstration_2018}. 
However, in case of the IST qubit circuit the phase particle stays trapped within its local minimum due to the high barrier formed by the large Josephson energy and the heavy mass (large shunt capacitance) of the phase particle. Only when the local minimum exceeds a critical value, which in our case is more than one $\Phi_0$ in flux bias or $\approx 15$\,GHz in energy from the ground state, a probabilistic tunneling event is triggered by vacuum or thermal fluctuations as shown in Fig.~\ref{fig:spec}a (top). At the base temperature we do not observe any such switching events on the time scale of hours for bias values close to $\varphi_\text{ext}=\pi$. This is an extreme case for the expected $T_1$ protection of the flux states in this limit \cite{lin_demonstration_2018} and points at interesting new qubit and quantum memory encodings that will need to rely on more advanced circuit control schemes. 
%  in In another words, the periodicity only shows itself in devices which allow for flux transition to occur. In IST qubit, beyond half of flux quantum the phase particle continues to exist on a meta stable state, unlocking a hidden part of the Hamiltonian in Eq.

In order to regain the flux periodicity of the spectrum of Eq.~\ref{eq:fluxonium_hamiltonian} we controllably increase the temperature of the device with a heater on the mixing chamber plate of the dilution refrigerator. The spectrum is stable without additional tunneling events up to around 100\,mK above which we observe a drastic increase of the number of switching events, as shown in Fig.~\ref{fig:spec}a. These random events add up to a consistent, smooth and periodic flux dependence at 250\,mK, as shown in Fig.~\ref{fig:spec}b.
% shows the 
%flux dependent 
%spectrum for a temperature of 250\,mK at which point it is smooth and fully predictable. 
This measurement probes the plasmon spectrum starting from a thermal mixture of all accessible qubit states 
%including different flux states
% represents a thermal mixtru
%spectrum, until a tunneling event transfers the phase particle to its global ground state. The tunneling event is followed by a jump in spectrum explaining both the non periodic behavior and discontinuities in the spectrum. Consequently, in spectroscopy shown in Fig. \ref{fig:spec}a counting the discontinuities gives a picture of number of tunneling events. Increasing the temperature of the environment shows no increase in number of tunneling events until $100$ mK where the discontinuities grow gradually with temperature to unveil a new periodic spectrum (Fig. \ref{fig:spec}b) at $250$ mK. 
where each parabola 
%in Fig. \ref{fig:spec}b 
represents the first plasmon transition of an individual well in the circuit potential. 
Point \textbf{A} in Fig.~\ref{fig:spec}b shows the flux sweet spot where one specific well is located at its minimum energy.
%, cf.~with Fig.~\ref{fig: concept}b. 
Point \textbf{B} on the other hand indicates the degeneracy point between two neighboring wells, and point \textbf{C}
% and therefore forms a critical anchor point of the half of flux quanta. The third interesting point in Fig. \ref{fig:spec}b is indicated by $c$ and 
shows a second order degeneracy between two next-neighbor wells. 
The amplitude of any parabola beyond half flux bias gradually vanishes, which indicates that the probability to find the system in the higher energy neighboring well is significantly lower than to find it in its global minimum well. 

Importantly, and different from other mechanisms that can induce uncontrolled flux discontinuities, such as when external flux vortices move in the vicinity of the rf-SQUID loop, we can also reconstruct a smooth spectrum from low temperature data. Combining a set of independent flux sweep measurements at the base temperature of $7$\,mK and combining them in one plot yields the data shown in Fig.~\ref{fig:spec}c and d for the resonator dispersive shift and the qubit frequency, respectively. 
%The Fig. \ref{fig:spec}b can also be obtained in $7$ mK by overlaying the extracted qubit frequencies from multiple sets of $7$ mK spectroscopy shown in Fig. \ref{fig:spec}a. The result of this technique is shown in Fig\ref{fig:spec}c where different colors represents different measurement sets. 
Using the periodicity found in Fig.~\ref{fig:spec}b, we solve the Hamiltonian in Eq.~\ref{eq:fluxonium_hamiltonian} numerically using the scQubits python library \cite{Groszkowski2021} to obtain the eigenenergies and fit (black lines) the characteristic energies and the qubit-resonator coupling of device B as listed in Tab.~\ref{tab:summary}. More details about the fitting procedure that also takes into account weak coupling to parasitic modes is found in Materials and Methods. 
%\hl{these numbers do not match the table: $E_J=40.3832$, $E_C=0.127$ and $E_L=0.72$ and $g_{\circ}=107.9$ MHz}.
%The black lines on Fig\ref{fig:spec}c shows the result of the fitting with $E_J=40.3832$, $E_C=0.127$ and $E_L=0.72$ in GHz units and a coupling strength of $g_{\circ}=107.9$ MHz. The full fitted parameters of the four IST qubit is provided in the Table.\ref{tab:summary}.

The observed phase tunneling physics and flux frustration highlights that the IST qubit can be considered a close relative of the phase qubit where the very-high linear inductance acts as a current bias for the Josephson junction while preserving the shape of the potential well and suppressing the band dispersion \cite{martinis2004superconducting}. The observed flux trapping is also related to Ref.~\cite{masluk_implementation_2012}, where the escape of the phase particle is observed in a device formed by two parallel Josephson chains coupled capacitively to a resonator, as well as to the hysteresis observed in rf-SQUID type Josephson parametric amplifiers \cite{pogorzalek_hysteretic_2017}. Nevertheless, we are not aware of any realizations of this physics in a superconducting qubit or any other non-distributed single junction device. 
%like the one present here. 

%Due to the high stability of the device over the full flux range we can 
Finally, we report the time-domain characterization of the plasmon qubit transition. 
%within half of flux quantum range and analyze the potential limitations. 
All $T_1$ measurements over the full flux range of device C are shown in Fig.~\ref{fig:coherence}a. 
%and very similar for the other two devices.
%and device A and B show a very similar behavior. 
The energy relaxation of the $|p1\rangle$ state is shown on a logarithmic scale, as obtained from $120$ individual $T_1$ measurement sweeps equally distributed over the full flux range. 
% where the dots represent the average and the bars the standard deviation. 
We find no sign of a double decay, which would indicate the presence of a relevant amount of quasi-particle induced loss \cite{pop_coherent_2014, grunhaupt_granular_2019, yan_flux_2016} and the histogram reveals a single peaked normal distribution (bottom inset). %due to the presence of a finite number of quasi particle tunneling events, which consequently implies that quasi particle tunneling is not the main channel of relaxation of the IST qubit. 

Based on Fermi's golden rule alone \cite{schoelkopf_qubits_2003}, we do not expect a $T_1$ dependence on the external flux 
% it is expected for the relaxation times to show no dependency with external flux 
since the transition frequency and its matrix element stays approximately constant over the entire flux range. Experimentally (Fig.~\ref{fig:coherence}a top inset) we observe a random variation around an otherwise constant mean of $T_1=15.5\,\mu s$, corresponding to an effective quality factor of $Q_q=0.67\times10^6$, on par with some of the best values in the literature \cite{nguyen_high-coherence_2019}.
%as  shown in the top inset of Fig.~\ref{fig:coherence}a.
The relatively high matrix element and transition frequency of the plasmon state render it susceptible to dielectric losses and the observed variation indicates possible two-level-system coupling~\cite{burnett_decoherence_2019}. 
%Dielectric surface losses the likely dominant factor 
%\hl{(did you check what is the Purcell limit taking into account the parasitic mode? would be nice to know what it is for the regular mode also, either via the single mode approximation or for the proper simulation)} 
% and that the main limitation is most likely to be caused by dielectric losses. 
%Assuming an effective quality factor of $Q_d=0.67\times10^6$ at the qubit transition frequency results on a $T_1$ limitation of $15.5\mu s$ as indicated by black dashed line in the top inset of the Fig.~\ref{fig:coherence}a. 
Material and design improvements based on a study of the participation ratios of the electric field distribution \cite{wang_surface_2015} and its interaction with the geometric superinductor could potentially overcome this limitation. 
%Finally, the lower inset of the Fig.~\ref{fig:coherence}a shows a normal distribution for relaxation times which further confirms that $T_1$ are not limited by flux noise.

%The relatively high matrix element and transition frequency of the plasmon state leaves the relaxation times of the qubit susceptible to dielectric losses. Using  Fermi's golden rule \cite{schoelkopf_qubits_2003} yields an effective mean quality factor of \hl{XYZ}, close to some of the best values in the literature \cite{nguyen_high-coherence_2019,smith_superconducting_2020}.

%The systematic flux noise protection present itself 
In Fig.~\ref{fig:coherence}b and c we compare the effect of flux noise protection for device C 
%with an 
%by showcasing the dephasing times variation with external flux for two samples, one with an i
with $E_L=1.6$\,GHz 
%(Sample A in Table.\ref{tab:summary}) 
and device A with $E_L=0.56$\,GHz
over the full flux range, respectively. 
%(sample C in Table.\ref{tab:summary}), within a half of flux quantum range. 
Device C shows a significant drop in measured $T_2$ times away from the flux sweet spot while device A with the three times higher inductance exhibits an approximately constant $T_2$ time over the full flux range. % as the derivative of the spectrum with respect to external flux grows larger. 

We model and fit (black dashed lines) the total decoherence rate with $\Gamma_{T_2}=\Gamma_{1/f}+\Gamma_\text{th}+1/(2T_1)$, where $\Gamma_{1/f}$ is due to flux noise and $\Gamma_\text{th}$ due to resonator photon shot noise.  %and $1/2T_1$ fix the dephasing limitation on the sweet spot. 
Dephasing due to $1/f$ flux noise can be expressed as 
$\Gamma_{1/f}=\sqrt{\gamma A_{\Phi}}\, |\pder[\omega_{p01}]{\Phi_\text{ext}}|$
and using Eq. \ref{eq:disper} we obtain
%and replacing the derivative in Eq. \ref{eq:fluxnoise} we have:
\begin{align}
T_{1/f}=\frac{1}{\Gamma_{1/f}}=\frac{\hbar\Phi_0(2E_J/E_L)^2}{4\pi\varphi_\text{ext}\sqrt{\gamma A_{\Phi}}\sqrt{8E_J E_C}},
\end{align}  
where $\sqrt{A_{\Phi}}\approx98\,\mu\Phi_0$ is the flux noise amplitude and $\gamma=\ln{\frac{f_u}{2\pi f_l}}\approx9.9$ represents the scaling parameter for the specific Ramsey sequence noise filter function with low and high frequency cutoffs $f_l= 250$\,mHz (inverse measurement time per data point) and $f_u=1/T_2=31$\,kHz \cite{Ithier2005,vepsalainen2021}.
%. based on a total measurement time defining the low frequency cutoff $f_l=1/(4\text{s}) = 250$/,mHz and a high frequency cut of $f_u=1/T_2=31$\,kHz due to the $T_1$ limit of $T_2$.
%. The $T_{1/f}$ contribution is depicted in Fig.~\ref{fig:coherence}b (dashed yellow line), using a fitted flux noise amplitude of $A_{\Phi}=98\,\mu\Phi_0$ and sacaling parameter of $\gamma=\ln{\frac{f_u}{2\pi f_l}}=9.9$ resulting from a total measurement time of $f_l=1/(4\text{s})=250mHz$ and a high frequency cut of $f_u=1/T_2=31$KHz\hl{(need to referenc:PHYSICAL REVIEW B 72, 134519 2005)}. 
%\hl{for both devices - it should be consistent - is it? why is this not in units of mu phi0? if it is we have phi0 squared  as unit though for phi times A}.  
This flux noise amplitude is found to be larger than the typical values reported in the literature, which we attribute to the large effective loop perimeter created by the geometric superinductance \cite{braumuller_characterizing_2020,peruzzo_geometric_2021}. It's contribution to the total dephasing is depicted in Fig.~\ref{fig:coherence}b (dashed yellow line).
%, larger than the values 

The flux independent thermal photon induced dephasing is calculated according to Ref.~\cite{rigetti_superconducting_2012}
%with an expression of the form:
%\begin{align}
%T_{th}=\frac{\kappa_L}{2}Re\qty[\sqrt{\qty(1+\frac{2i\chi}{\kappa_L})^2+\qty(\frac{8i\chi n_{th}}{\kappa_L})}-1]
%\end{align}  
%Where the $\kappa_L$ is the line with of the cavity resonance, $\chi$ is the dispersive shift of the qubit excitation on the resonator and the $n_{th}$ represents the thermal occupation in the resonator. In the case of the device C with $E_L=1.6$ GHz the resonator line width is $\kappa_L=809 KHz$ and the dispersive shift is measured to be $\chi=233$ KHz. 
and shown together with the measured $1/(2~T_1)$ limit in Fig.~\ref{fig:coherence}b and c (cyan dashed lines). From the fit we obtain
%shows the thermal photon limitation, 
%assuming 
%n_th=0.008 device C
%n_th=0.028 device A
a thermal resonator occupation of $n_\text{th}=0.009$ for device C shown in panel b and $n_\text{th}=0.028$ for device A shown in panel c. The difference could be explained partly by the fact that the resonator of device A is coupled stronger to external drive and readout line but we note that it's coherence might also be limited by a different flux independent dephasing mechanism. 
%(I would swap the temperatures to the main text and the occupations to the figure caption).}

The effective dephasing model (black dashed line) agrees well with the measured $T_2$ times of device C shown in Fig.~\ref{fig:coherence}b. Devices A and B have the largest inductance and exhibit a much larger ratio $E_J/E_L \approx 53$, which results in a drastically reduced flux dispersion. While the $T_2$ data shown in Fig.~\ref{fig:coherence}b fluctuates as a function of flux we do not observe a systematic reduction of $T_2$ up to $\varphi_\text{ext}=\pi$. In case of device B (not shown) we observe a larger variation but the maximum $T_2 \approx 28.5~\mu$s is measured at $\varphi_\text{ext}=\pi/2$. Given the high flux noise amplitude these results represent a new level of dephasing protection in a flux tunable device.

%for device A shown in Fig.~\ref{fig:coherence}b. 
%and the smallest flux dispersion of only \hl{5\,MHz (how much exactly?)}.
%waveguide. , together with $\frac{1}{2T_1}$ limitation on the dephasing. 
%Finally, the black dashed line is the effective limitation on dephasing which shows a good agreement with measured $T_2$ times. In the case of the device A with an inductive energy of $E_L=0.6$ GHz (strong protection), the Eq. \ref{eq:fluxnoise} no longer fits and $T_2$ values show no variation with external flux(as shown in bottom picture in Fig\ref{fig:coherence}b). This is due to high $E_J/E_L$ ratio suppressing the dispersion down to only a $5$ MHz. In this device the only limitation comes from thermal photon noise and $1/2T_1$ rates.

\section{Discussion}
In summary, we have theoretically and experimentally introduced a new parameter regime for superconducting qubits: the inductively shunted transmon (IST), which is characterized by very large $E_J/E_C\sim100$ and $E_J/E_L\sim50$ energy ratios. While the transmon is derived from the Cooper pair box circuit, the IST qubit is derived from the rf-SQUID circuit, closest to an ultra-heavy fluxonium or an ultra-high inductance flux qubit. Nevertheless, we show that the properties of the low lying plasmon spectrum closely resemble those of the transmon but now with flux tunability and without charge dispersion. On a conceptual level its potential and wavefunctions are continuous and extended in contrast to the periodic potential of the transmon. As a hallmark of this new regime we observe stable non-decaying fluxon states and thermally assisted quantum tunneling in a single-mode superconducting qubit. 

The present work focusses on the properties of the plasmon encoding and we identified the characteristic $E_J/E_L$ ratio as the relevant parameter to carefully control the band dispersion and the resulting flux noise sensitivity of the device. With a demonstrated flux dispersion of only $5.1$\,MHz over a full flux quantum it is significantly less noise sensitive compared to the high impedance approach investigated to date \cite{pechenezhskiy_superconducting_2020, peruzzo_geometric_2021, Smith2020}. Combined with a lower flux noise amplitude inductor and lower TLS density capacitor materials, as well as an improved geometry to reduce surface loss participation, the IST concept opens a new path forward to introduce in-situ fine tuning of the transmon frequency without sacrificing protection against flux noise. 
%of the single junction transmon. 

In a regular transmon qubit, strong excitations, useful e.g.~for high fidelity qubit readout or stabilized bosonic qubit implementations \cite{grimm_stabilization_2020}, can easily exceed the weakly anharmonic ladder of confined states within the cosine potential. This can cause instabilities in the average number of excitations \cite{lescanne_escape_2019} and lead to excitations out of the computational basis via non energy conserving terms of the Jaynes-Cummings Hamiltonian \cite{sank_measurement-induced_2016}. The parabolic and non-periodic confinement of more moderate inductance value IST qubits based on linear geometric inductors is expected to better confine higher energy states and might be able to avoid such leakage \cite{verney_strongly_2018}. 
%This could help to avoid 
%It has been shown that high photon numbers in the cavity can excite a transmon qubits   It is shown theoretically that the IST qubit can potentially solve the instability since its parabolic potential confines all the higher energy states \cite{verney_strongly_2018}.

The fluxonium qubit platform has recently been identified as an alternative way forward to scaling up superconducting qubit processors \cite{Bao2021,Nguyen2022} due to promising coherence times, higher design flexibility and anharmonicity. The use of geometric inductors could offer advantages for the reproducibility of $E_L$ \cite{peruzzo_geometric_2021} and the current work shows that one of its major drawbacks, i.e.~an enhanced flux noise amplitude, could in principle be mitigated with a noise-insensitive design. 
%Also the behavior of the qubit while subjected to high power microwave radiation with resonator frequency is now actively under study. 

Flux qubit encoding in the IST limit presents some challenges due to the excessively low fluxon transition matrix elements on the order of $10^{-13}$ - the reason for the observed protection against energy relaxation from one flux well to another. Early results indicate however that high fidelity excited state preparation is possible, which enables careful studies of the time domain tunneling physics and the fluxon lifetimes. In addition, real-time control of the qubit characteristic energies such as the tunneling barrier $E_J$ might open a way for full qubit control \cite{Zhang2021}, as required to characterize the fluxon coherence, - a promising route towards new decay-protected qubit encoding schemes. 

Full control over both, the plasmon and fluxon qubit encoding could lead to interesting hybrid applications in non-adiabatically driven or dynamically controlled qubit circuits that intrinsically combine fast gates with memory elements. On a more fundamental level it might offer new capabilities to study quantum tunneling \cite{ramos2020b} in dynamically controlled potentials and our implementation based on a $\sim 14$\,mm long SQUID wire might revive the quest for pushing the macroscopicity in superconducting quantum circuits \cite{Leggett1980,friedman2000,Korsbakken2010,Nimmrichter2013}.

The data and code used to produce the figures in this manuscript will be made available at Zenodo. 

\section*{Acknowledgments}
The authors thank J. Koch for discussions and support with the scQubits python package, S. Barzanjeh and G. Arnold for theory, E. Redchenko, S. Pepic, the MIBA workshop and the IST nanofabrication facility for technical contributions, as well as L. Drmic and P. Zielinski for software development. This work was supported by a NOMIS foundation research grant, the Austrian Science Fund (FWF) through BeyondC (F7105), and IST Austria. 
%M.P. is the recipient of a P\"ottinger scholarship at IST Austria. E.R. is the recipient of a DOC fellowship of the Austrian Academy of Sciences at IST Austria.

\section*{Methods}

\subsection{Perturbation theory}
\begin{figure}[t]
\centering
\includegraphics[width=0.8\columnwidth]{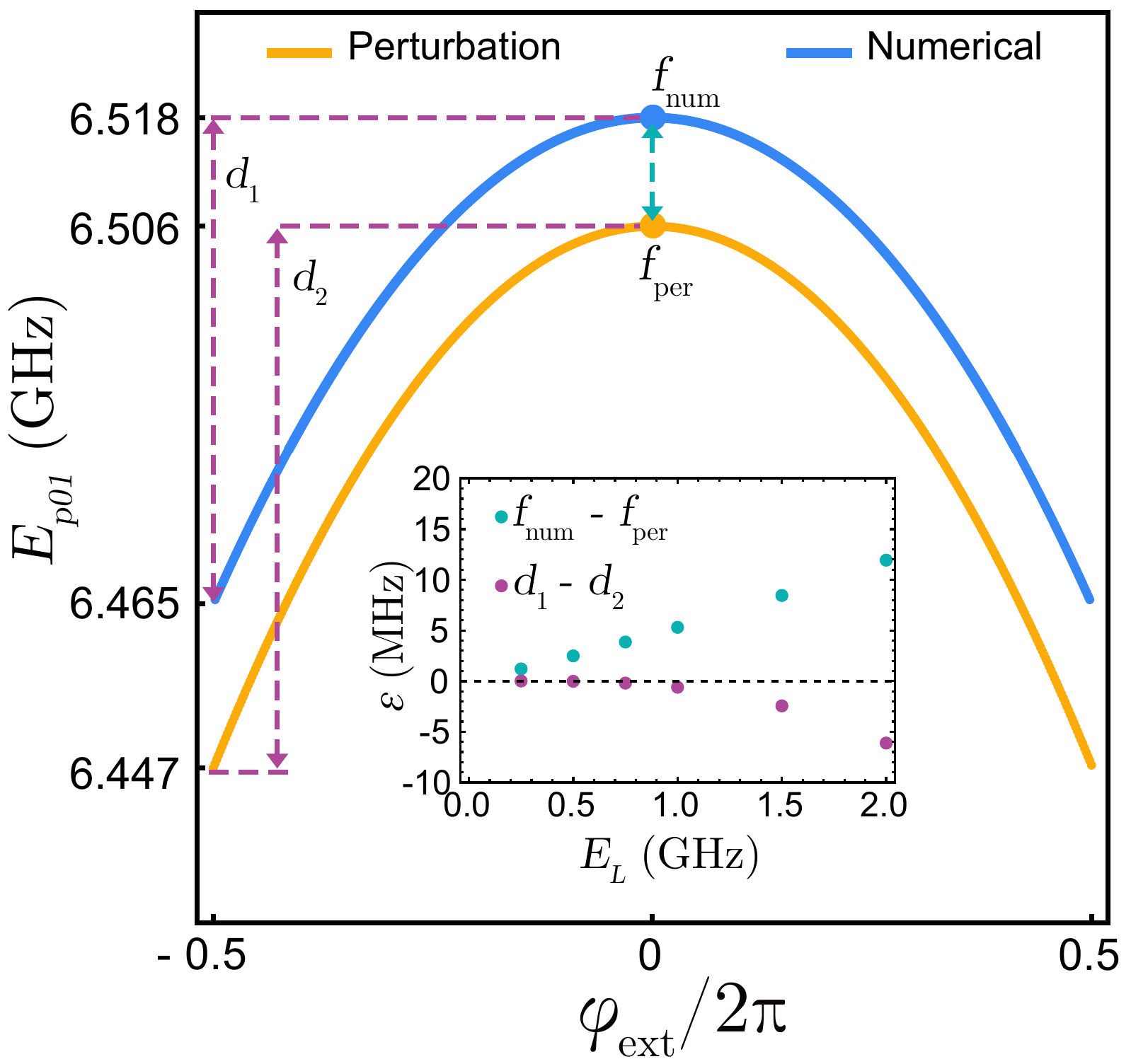}
\caption{Typical ground to excited state transition frequency of the IST qubit with $E_J/h=35$, $E_C/h=0.15$ and $E_L/h=2$\,GHz, calculated numerically (blue) and predicted using perturbation theory (yellow). The analytic solution shows a deviation $\epsilon$ in predicting the frequency (green arrow) and dispersion (purple arrows). The inset shows how these errors scale with the perturbation strength $E_L$. As the inductive energy decreases the error in both frequency and dispersion converge to the numerical results.} 
\label{fig:purturbation}
\end{figure}

We use perturbation theory to investigate the degree of protection characterized by the $E_J/E_L$ ratio. We split the Hamiltonian in Eq.~\ref{eq:fluxonium_hamiltonian} into the transmon part and an inductive part that we treat as a perturbation
\begin{align}
&\hat{H}=\hat{H}_{\text{trans}}+\hat{H}_{\text{per}},\nonumber\\ 
&\hat{H}_{\text{trans}}=E_C\hat{n}^2-E_J\cos{\hat{\phi}},\\
&\hat{H}_{\text{per}}=\frac{1}{2}E_L(\Hat{\phi}+\varphi_\text{ext})^2.\nonumber
\quad
\end{align}
Then, by following the formalism of perturbation theory \cite{griffiths2018introduction}, we can calculate the corrections to transmon eigenenergies up to second order
\begin{align}
E_m=E^{(0)}_m+E^{(1)}_m+E^{(2)}_m,
\end{align}
where $E^{(0)}_m$ being the eigenenergies of the transmon derived from Mathieu functions \cite{koch_charge-insensitive_2007} and $E^{(1)}_m$ and $E^{(2)}_m$ representing the first and second order energy correction respectively. These values can be calculated using
\begin{align}\label{eq:corr}
&E^{(1)}_m=\mel{m}{\hat{H}_{\text{per}}}{m},\nonumber\\
&E^{(2)}_m=\sum_{m\neq n}\frac{\abs{\mel{n}{\hat{H}_{\text{per}}}{m}}^2}{E^{(0)}_m-E^{(0)}_n}.
\end{align}
Now, to find an expression for $E^{(1)}_m$ and $E^{(2)}_m$, we employ the second quantization formalism of the flux operator
\begin{align}\label{eq:secquan}
&\hat{\phi}=\left(\frac{2E_C}{E_J}\right)^{\frac{1}{4}}\left(\hat{a}^\dagger+\hat{a}\right),\\
&\hat{a}^{\dagger}|n\rangle=\sqrt{n+1}|n+1\rangle,\\
&\hat{a}|n\rangle=\sqrt{n}|n-1\rangle,
\end{align}
with $\hat{a}^\dagger$ and $\hat{a}$ being the raising and lowering operators for the transmon and $n$ is the transmon qubit state number. Substituting Eq.~\ref{eq:secquan} in the first order energy correction term shown in Eq.~\ref{eq:corr}, we obtain
\begin{align}\label{eq:firstcorr}
    E_{n}^{(1)}=\frac{1}{2\hbar}\left(2 n+1\right) E_{L}\left(\frac{2 E_{C}}{E_{J}}\right)^{\frac{1}{2}}+\frac{1}{2\hbar} E_{L}\varphi_\text{ext}^{2}.
\end{align}
In the first order energy correction shown in Eq.~\ref{eq:firstcorr}, the external flux appears with a prefactor that is not state number dependent. Consequently, the calculated transition energies using only the first order correction will not reflect the external flux effect on the transition. Therefore, we extend the calculations to second order terms. The expression for ground and excited second order energy corrections are 
\begin{align}
&E_{0}^{(2)}=\frac{1}{4\hbar} E_{L}^{2}\left[\frac{2\left(\frac{2 E_{C}}{E_{J}}\right)}{E_{0}^{(0)}-E_{2}^{(0)}}+\frac{4\left(\frac{2 E_{C}}{E_{J}}\right)^{\frac{1}{2}} \varphi_\text{ext}^{2}}{E_{0}^{(0)}-E_{1}^{(0)}}\right],\nonumber \\
&E_{1}^{(2)}=\frac{1}{4\hbar} E_{L}^{2}\left(\frac{2 E_{C}}{E_{J}}\right)\left[\frac{6}{E_{1}^{(0)}-E_{3}^{(0)}}\right]+\nonumber\\ 
&\frac{1}{4\hbar} E_{L}^{2}\left(\frac{2 E_{C}}{E_{J}}\right)^{\frac{1}{2}}\left[\frac{8 \varphi_\text{ext}^{2}}{E_{1}^{(0)}-E_{2}^{(0)}}+ \frac{4 \varphi_\text{ext}^{2}}{E_{1}^{(0)}-E_{0}^{(0)}}\right],
\end{align}
while the general expression for $n\gg2$ is
\begin{align}
&E_{n \geq 2}^{(2)}=\frac{E_{L}^{2}}{\hbar}\left(\frac{ E_{C}}{2E_{J}}\right)\left[\frac{(n+1)(n+2)}{E_{n}^{(0)}-E_{n+2}^{(0)}}+\frac{(n)(n-1)}{E_{n}^{(0)}-E_{n-2}^{(0)}}\right]+\nonumber \\
&\frac{E_{L}^{2}}{\hbar}\left(\frac{2 E_{C}}{E_{J}}\right)^{\frac{1}{2}}\varphi_\text{ext}^{2}\left[\frac{ (n+1)}{E_{n}^{(0)}-E_{n+1}^{(0)}}+\frac{n}{E_{n}^{(0)}-E_{n-1}^{(0)}}\right].
\end{align}
The second order energy correction terms are also dependent to $\varphi_\text{ext}$ but with a prefactor determined by the transmon state number. In Fig.~\ref{fig:purturbation} the result of a numerical calculation of a typical IST qubit's first transition frequency with $E_J/h=35, E_C/h=0.15, E_L/h=2$ (all in GHz) is plotted against a prediction from perturbation theory. The analytic solution has inaccuracies both in frequency and dispersion but for small $E_L$ the results converge to the exact numerical solution as shown in the inset of Fig.~\ref{fig:purturbation}).

\par Finally, to arrive at Eq.~\ref{eq:disper}, we calculate the derivative of the first transition
\begin{align}\label{eq:derivative}
    &\pder[\omega_{p01}]{\varphi_\text{ext}}=\nonumber\\
    &4\frac{E_{L}^{2}}{\hbar} \varphi_\text{ext}\left(\frac{2 E_{c}}{E_{J}}\right)^{\frac{1}{2}}\left[\frac{1}{E_{1}^{(0)}-E_{2}^{(0)}}+\frac{1}{E_{1}^{(0)}-E_{0}^{(0)}}\right].
\end{align}
Now, using the approximation of $E_{1}^{(0)}-E_{2}^{(0)}=-\sqrt{8E_J E_C}-2E_C$ and $E_{1}^{(0)}-E_{0}^{(0)}=\sqrt{8E_J E_C}-E_C$ in the limit of high $E_J/E_C$, we can further simplify the expression in Eq.~\ref{eq:derivative} to
\begin{align}
    &\pder[\omega_{p01}]{\varphi_\text{ext}}=4\frac{ E_{L}^{2}}{\hbar}\varphi_\text{ext}\left(\frac{2 E_{C}}{E_{J}}\right)^{\frac{1}{2}}\times\nonumber\\
    &\left[\frac{-E_{C}}{\left(\sqrt{8 E_{J} E_{C}}-2 E_{C}\right)\left(\sqrt{8 E_{J} E_{C}}-E_{C}\right)}\right],
\end{align}
which, by ignoring the $E_C$ terms in the denominator in comparison with plasmon frequency $\sqrt{8E_J E_C}$, can be further simplified to:
\begin{align}
    \pder[\omega_{p01}]{\varphi_\text{ext}}=-\frac{\sqrt{8E_J E_C}}{\hbar(2E_J/E_L)^2}\varphi_\text{ext}.
\end{align} 
It is important to mention that this theoretical description actually models a periodic parabolic potential as the perturbation and only represents our system (the IST qubit) in a local sense, i.e.~within the first flux quantum and for $\phi\in(-\pi,\pi)$. Without taking this into consideration the wavefunctions obtained from perturbation theory will not be periodic and therefore contradict the periodic transmon wavefunctions. Beyond the first flux quantum perturbation theory fails to predict the IST qubit properties simply because its potential is different than that of the IST qubit.  

\begin{figure*}[t]
\centering
\includegraphics[width=2.1\columnwidth]{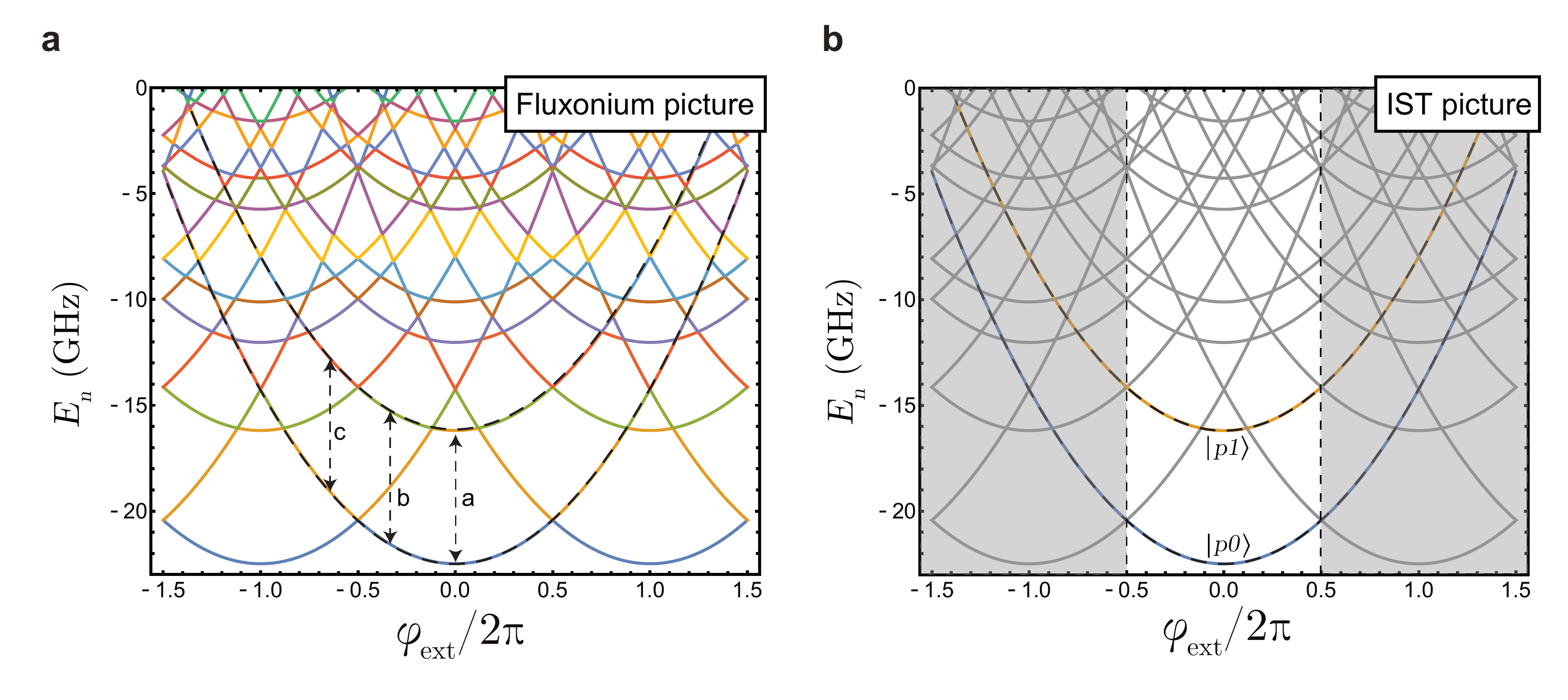}
\caption{Numerically calculated eigenenergies of device B in Table.~\ref{tab:summary} using the scQubits library. a, shows eigenenergies colorcoded by their respective state number $m$. The black dotted line guides the eye on the two lowest states located in one specific potential well and is measured experimentally as the first transition. Position $a$ labels the transition correctly as $\ket{p0}\rightarrow\ket{p1}$, while in position $b$ and $c$ the desired transition refers to $\ket{p0}\rightarrow\ket{p2}$ and $\ket{p1}\rightarrow\ket{p3}$ respectively. The reason for this is that in fluxonium the flux transitions are allowed and the system is always found in its global minimum while in the IST qubit the phase particle is trapped in a stable flux configuration and stays in one specific well. b, shows a quadratic fit to the ground to first excited state transition of the IST qubit. In the IST regime all the eigenenergies of any state are represented by a parabola vs. flux which makes the fitting process significantly easier.}       
\label{fig:fit}
\end{figure*}

\subsection{Device fabrication}
The fabrication of the IST qubit starts with cleaning a $10\times10~\text{mm}^2$ high resistivity silicon chip using an $O_2$ plasma asher followed by a buffered hydrofluoric acid dip, sonication in acetone for 10 minutes at a temperature of $50^{\circ}\text{C}$ and a final rinse with isopropanol (IPA). To form the alignment markers and identifiers, the chip was covered with AR-P $6200$ (CSAR 62) resist and patterned to dry etch with an Oxford ICP machine. After ICP dry etching, the chip is solvent cleaned with N-Methylpyrrolidone (NMP) followed by an acetone and IPA rinse. 

In the next layer, the CSAR 62 was also used to lift-off the first layer of aluminum with a thickness of $100$\,nm forming the crosswire which provides access to the inner pad of the geometric superinductor in the subsequent layers. To protect the crosswire and to shape the air-bridges as smooth arcs (see Fig.~\ref{fig:device}), the CSAR 62 was patterned and reflowed at $180^{\circ}\text{C}$. 
%to give the curvature on the aluminum bridges. 
Then $150$\,nm of aluminum was evaporated while the chip was tilted at $10$ degrees and the sample holder was in rotation. The tilt and rotation helps to cover the resist in the previous step uniformly. subsequently,  aluminum was dry etched using a calibrated mixture of BCl$_3$ - Cl$_2$ gases to form the geometric superinductor and capacitive antenna pads. 

Next, the sample was covered with a double layer of MMA/PMMA resist to pattern the Dolan bridge \cite{dolan_offset_1977} for Josephson-junction shadow evaporation. Before junction evaporation starts, we use an in-situ gentle argon ion milling process to clean the surface of the silicon of possible residues left from resist development. The process uses a $250$\,V as the acceleration voltage and a current of $10$\,mA with an argon flow of 4.5\,sccm. Then the Josephson junction was fabricated by first evaporating a $60$\,nm aluminum as the base electrode of the junction followed by a calibrated static oxidation with oxygen at a pressure of $5$\,mbar and $5$\,minutes to reach the Josephson energy of $35$\,GHz (with an area of $250\times250\, \text{ nm}^2$), and finally the counter electrode of the junction was evaporated with a thickness of $120$\,nm. 

The final layer of the device connects all the previous layers with a suitable patch. To remove the aluminum oxide we use in situ argon ion milling with more aggressive parameters ($400\,\text{V}$, $21\,\text{mA}$ and $4.5\,\text{sccm}$ argon flow) for $5$ minutes. Since this layer has to cover all the previous layers, a $300$\,nm thick aluminum film was used. A double layer PMMA resist with a thickness of approximately $1\,\mu\text{m}$ was used to assist the final lift-off process. Finally, the chip was covered with a S1805 photo resist and a UV tape to be diced into three $10\times2.5\,\text{mm}^2$ pieces.      

\subsection{ Fitting procedure}
To fit the spectroscopy data shown in Fig.~\ref{fig:spec}, we numerically solved (scQubits library) a fluxonium in the IST regime coupled to a resonator and extracted the einenenergies as shown in Fig.~\ref{fig:fit}a. At zero flux bias, the first transition is labeled correctly by $\ket{p0}\rightarrow\ket{p1}$ (position $a$ in Fig.~\ref{fig:fit}a), however, since the flux transition is not allowed in IST qubit and the phase particle is trapped within one well, the experimentally measured first transition at position $b$ refers to $\ket{p0}\rightarrow\ket{p2}$ while the same transition at position $c$ refers to $\ket{p1}\rightarrow\ket{p3}$. The constant change in state numbers makes the fitting of the experimental data challenging.    

In Fig.~\ref{fig:fit}a, one may notice that the eigenenergies for any state is a quadratic function of external flux. The intuition for this observation is provided in \cite{koch_charging_2009}, where applying a transformation to the Hamiltonian of Eq.~\ref{eq:fluxonium_hamiltonian} into the Bloch wave basis and in the limit of high $E_J/E_C$ ratio, block diagonalizes the Hamiltonian into separate effective Hamiltonians expressed as
\begin{align}\label{eq:dual}
H^{(s)}=\frac{E_{L}}{2}\left(i \frac{d}{d p}+\frac{2 \pi \Phi}{\Phi_{0}}\right)^{2}+\varepsilon_{s}(p),
\end{align}
where $p$ is the quasi momentum and $s$ is the band index of the corresponding CPB Hamiltonian \cite{koch_charging_2009}. In the IST qubit case, where the $E_J/E_C$ is in the transmon limit the $\varepsilon_{s}(p)$ can be written as \cite{koch_charge-insensitive_2007}
\begin{align}
\varepsilon_{s}\left(p\right) \simeq E_{s}\left(p=1 / 4\right)-\frac{\epsilon_{s}}{2} \cos \left(2 \pi p\right).
\end{align}
Deep in the transmon limit, $\epsilon_{s}$ is exponentially suppressed \cite{koch_charge-insensitive_2007} and therefore the Hamiltonian in Eq.~\ref{eq:dual} represents the IST qubit as a free particle with a quadratic dispersion with external flux. In Fig.~\ref{fig:fit}b, the lowest state of a corresponding well is identified by simply fitting a quadratic function. Therefore the first transition was calculated by subtracting the two fitted parabola and fitted to the experimental data presented in Fig.~\ref{fig:spec}. 

It is important to note that the correct qubit fit parameters for devices A and B were obtained by also including the coil parasitic modes found close to the first transition frequency. In case of device A, $\nu_{p01}=6.1222$\,GHz and the parasitic mode $\nu_p=6.1890$\,GHz with a coupling of $15.9$\,MHz, while for device B the first qubit transition was at $6.296$\,GHz and the parasitic mode located at $6.13$\,GHz with a coupling of $22$\,MHz. In case of device C the parasitic mode was at sufficiently high frequency to not affect the fitting procedure.

\bibliography{ISPO}
\end{document}